\documentclass[manuscript]{aastex}

\slugcomment{Submitted to the Astronomical Journal (revised version)}

\shorttitle{New distant companions to known nearby stars. I.}
\shortauthors{Lepine et al.}

\begin{document}

\title{New distant companions to known nearby stars: I. GJ 4047B,
GJ 718B, GJ 747.2C, GJ 4100B, and GJ 4153B.\altaffilmark{1}}

\author{S\'ebastien L\'epine\altaffilmark{2} and Michael M. Shara}
\affil{Department of Astrophysics, Division of Physical Sciences,
American Museum of Natural History, Central Park West at 79th Street,
New York, NY 10024, USA}

\and

\author{R. Michael Rich}
\affil{Department of Physics and Astronomy, University of California
at Los Angeles, Los Angeles, CA 90095, USA}

\altaffiltext{1}{Based on data mining of the Digitized Sky Survey,
developed and operated by the Catalogs and Surveys Branch of the Space
Telescope Science Institute, Baltimore, USA.}
\altaffiltext{2}{Visiting Astronomer, Lick Observatory}

\begin{abstract}
In an ongoing survey for high proper motion stars at low galactic
latitudes ($-25^{\circ}<b<25^{\circ}$), we have identified 5
previously uncatalogued common proper motion companions to stars
listed in the {\it Preliminary Version of the Third Catalogue of
Nearby Stars} (GJ stars). For each system, the relative proper motion
between the components is less than 5\% of the common proper motion of
the pair. Spectra of the pairs have been obtained at the Lick 3-m
Shane Telescope, confirming that the systems are indeed wide
separation binaries. The systems are classified as follows: GJ 4047AB
= K5 V + M5 V, GJ 718AB = K5 V + M4.5 V, GJ 747.2ABC = (K7 V + K7 V) +
M4 Ve, GJ 4100AB = M1 V + M4.5 Ve, and GJ 4153AB = M0.5 V + M3.5 V. The
total area surveyed contains 346 Gliese stars, which suggests that
$\approx 1.5\%$ of the stars listed in the {\it Preliminary Version of
the Third Catalogue of Nearby Stars} have unrecognized proper motion
companions. We predict that $\approx$15-50 more new distant companions
to GJ stars will be discovered in the Digitized Sky Survey.
\end{abstract}

\keywords{Stars: low-mass, brown dwarfs --- Binaries: visual ---
Stars: fundamental parameters}

\section{Introduction}

The current census of stars in the solar neighborhood is believed to
be significantly incomplete, even within only 10pc of the sun (Henry
{\it et al.} 1997, Delfosse {\it et al.} 2000). The ongoing search for
nearby stars is a challenge because the vast majority of stellar
objects in the Galaxy consist of intrinsically faint M dwarf stars,
whose absolute visual magnitudes are in the $10<M_V<20$ range. Not only
are there several stellar systems waiting to be discovered, but it is
very likely that a significant number of known nearby systems are yet
to be recognized as multiple stars. A census of the stars in the
immediate vicinity ($d<5$pc) of the sun reveals that more than half of
the stellar bodies are members of double and triple stellar systems
(van de Kamp 1971). On the other hand, the {\it Third Catalog of
Nearby Stars} by Gliese \& Jahreiss 1991 (hereafter GJ), lists only
1031 bodies in multiple systems out of a total of 3803 stars.

Multiplicity among nearby M dwarf stars has been discussed and
estimated by Fischer \& Marcy (1992). Methods for the identification
of companions depend on the separation (semi-major axis $a$) between
the components. For stars in the solar neighborhood, very close
companions $a<1$ AU can generally be identified as spectroscopic
binaries. Systems with intermediate separations (1 AU$<a<$25 AU) are
best found with high-resolution imaging techniques. Nearby wide
binaries, on the other hand, are well resolved on the sky, even on old
photographic plates, and the challenge is to distinguish them from
background stars. The main method to identify the components of nearby
wide binaries is through their common proper motion.

While close and intermediate separation binaries require intensive
observational efforts, wide binaries can be identified using available
all-sky surveys, provided they are sensitive enough to detect the
companion as a high proper motion object. A list of wide binaries has
been compiled by Poveda {\it et al.} 1994 (hereafter PHACL). Expanding
on the GJ catalog, they searched for unaccounted common proper motion
companions to GJ stars within 22.5pc of the sun by correlating the GJ
stars with entries in the {\it catalog of Double Stars with Common
Proper Motion} (LDS catalog, Luyten 1987) and in the {\it New Luyten
Catalogue of stars with proper motions larger than two tenths of an
arcsecond} (NLTT catalog, Luyten 1980). They found 18 more systems
overlooked by GJ. The compilation given in PHACL includes 305 wide
binaries, 26 triples, and 3 quadruples, all systems with a semi-major
axis $a>25$AU, i.e. with an angular separation exceeding 1$\arcsec$. The
fact that PHACL reports 18 companions not listed in GJ strongly
suggest that there are many more companions to be found, because {\em
both the NLTT and LDS catalogs are themselves significantly
incomplete}.

Two recent studies have provided a striking demonstration of the
effect of incompleteness of high proper motion star catalogs on the
current census of wide binary systems. McCarthy, Zuckerman, \& Becklin
(2001) have discovered one previously unrecognized, faint common
proper motion companion to the star G144-016 (GJ 4153). They
serendipitously found the faint common proper motion companion
(located 1$\arcmin$ from G144-016) by comparing their finding charts drawn
from first and second epoch Digitized Sky Survey (DSS) images. In
another survey for substellar companions to nearby stars, Kirkpatrick
{\it et al.} (2001) report the discovery of three faint companions to
known nearby stars based on their IR colors as obtained from the 2MASS
survey. One of the secondaries (G216-7 B) was also identified as a
previously unrecognized faint proper motion companion, visible on the
first and second epoch DSS images. These two studies suggest that a
systematic search for proper motion companions on Palomar sky survey
plates should reveal several more distant companions to known nearby
stars.

We have initiated a systematic, automated search for high proper
motion stars using the DSS (L\'epine {\it et al.} 2002, in
preparation). We are particularly interested in the fields at low
galactic latitudes, where the NLTT catalog is clearly deficient
($\approx1$ star per square degree at $-10^{\circ}<b<+10^{\circ}$,
compared with $\approx3$ stars per square degree at
$b>+80^{\circ}$). We are using the SUPERBLINK software, developed by
SL, which works as an automated blink comparator. We have so far
discovered several thousand stars that are well within the limits of
the NLTT catalog ($\mu>0.18\arcsec$ yr$^{-1}$, $R<19$) but have been
missed by Luyten for a variety of reasons. We suspect that most of the
stars have been missed because of the obvious difficulties involved
when visually searching for moving objects in very crowded fields with
the blink comparator. A computer-based search proves to be much more
powerful and efficient.

By comparing the list of new high proper motion stars found by our
software with objects listed in the GJ catalog, we have discovered 5
common proper motion companions to GJ objects previously listed as
single stars. In this paper, we present spectroscopic observations of
those newly identified wide multiples. One of the companions, GJ
4153B, has already been discovered serendipitously by McCarthy,
Zuckermann, \& Becklin (2001). The other four common proper motion
companions are being cited here for the first time. This paper
presents spectroscopic follow-up of the five common proper motion pairs,
which confirms that they are physically related, low-mass, M dwarf
companions to the known GJ stars. We discuss the potential for new
discoveries in the conclusion.

\section{Identification Method}

We are conducting a systematic, automated survey for high proper
motion stars at low galactic latitudes using the SUPERBLINK
software. Using a differential technique, SUPERBLINK automatically
searches for and identifies variable and moving objects on sub-images
of the first and second Palomar Sky survey retrieved from the DSS. The
code identifies stars with proper motions $\mu\gtrsim100$ mas
yr$^{-1}$ and measures their proper motions with an accuracy of
$\approx5$ mas yr$^{-1}$. Each candidate high proper motion star is
verified visually by blinking $2\arcmin\times2\arcmin$ sub-images
centered on the candidate; bogus detections from e.g. plate artifacts,
dust specks, scanning errors, are thus eliminated. As a preliminary
study, we have investigated an area of the sky delimited by
$DEC>-2.5^{\circ}$, $16h<RA<22h$, and covering the low galactic
latitudes region ($-25^{\circ}<b<+25^{\circ}$). Within these limits,
our SUPERBLINK software has identified 18,953 stars with
$\mu>0.10\arcsec$ yr$^{-1}$, all of which have been confirmed by
visually blinking the candidates' images. We have found a total of
4,287 stars with $\mu>0.18\arcsec$ yr$^{-1}$, of which 2,107 are
``new'' stars not listed in the NLTT catalog but within its limits.

The area under study covers an area $\approx4500$ square degrees, and
contains a total of 346 stars listed in the GJ catalog. We used the
VizieR web interface at the {\it Centre de Donn\'ees Astronomiques de
Strasbourg} to correlate all of our stars with $\mu>0.10\arcsec$
yr$^{-1}$ with the GJ catalog of nearby stars. We looked for any
object in our catalog located within $1.5\arcmin$ of a known nearby
star, and verified whether the object had a proper motion similar in
magnitude and orientation as the proper motion of the GJ star. We
identified 5 stars meeting this criterion which were {\em not} already
listed in the GJ catalog. Finder charts for the five pairs are shown
in Figures 1-5. In each case we show the POSS-I image to the left
(epoch$\approx$1950) and the POSS-II image to the right
(epoch$\approx$1990).

One of the companions we found is GJ 4153B, first identified by
McCarthy, Zuckermann, \& Becklin (2001). We verified with the SIMBAD
Astronomical Database and confirmed that the other 4 companions had
never been cited in the literature before. All 5 stars were selected
for follow-up spectroscopic observations, since McCarthy, Zuckermann,
\& Becklin (2001) did not publish a spectrum for GJ 4153B. 

\section{Spectroscopic Observations}

Spectra of the five systems were obtained at the Lick Observatory with
the KAST spectrograph mounted on the 3.0-m Shane Telescope, under
photometric conditions on the night of July 23-24, 2001 (heliocentric
Julian date 2,452,114). Longslit spectra were obtained with the stars
centered on a 2.5$\arcsec$ wide slit. The instrument rotator was used, and
set for each object such that the slit was oriented at the parallactic
angle in order to avoid slit losses due to atmospheric differential
refraction (Filippenko 1982). Four spectroscopic standards were
observed during the run (Feige 110, PG 1545+035, PG1708+602, and Wolf
1346, see Massey \& Gronwall 1990), and were used for calibration and
as templates for the removal of telluric lines. In order to correct
for the fringing on the CCD, which is significant on the red channel
of the KAST spectrograph redward of 7500 \AA, we obtained separate
dome flats for each target at the same telescope orientations
(azimuth-altitude) as the observations.

Reduction was carried out with IRAF, using the standard procedure for
the reduction of longslit spectra (DOSLIT), including calibration and
removal of the telluric lines. The resulting spectra are presented for
each system in Figures 1-5, with both the primary and the secondary
plotted on the same graph {\em on a logarithmic scale} to emphasize the
difference in the monochromatic fluxes.

Our spectral classification is based on the strength of the TiO5,
CaH2, and CaH3 spectral indices as defined and calibrated in Reid {\it
et al.} (1995) and Gizis (1997). We have also used the CaH1/TiO5,
CaH2/TiO5, and CaH3/TiO5 ratios (see Gizis 1997) to determine whether
the stars are dwarfs or subdwarfs; the M stars are all clearly dwarfs
under this system. The values of the TiO5, CaH1, CaH2, CaH3 indices
are listed in Table 2 along with the estimated spectral types. The
estimated spectral types are the average spectral types obtained from
the Reid {\it et al.} (1995) TiO5 index calibration and the Gizis
(1997) TiO5, CaH2, and CaH3 indices calibrations; the accuracy is half
a spectral type.

\section{Discussion}

\subsection{{\it GJ 4047AB}}

GJ 4047 is the bright high proper motion star known as Ross 706. It is
the HIPPARCOS star HIPP-89656, a $V=9.710$ mag star with a proper
motion $\mu=345.8$ mas yr$^{-1}$ and parallax $px=31.65\pm1.47$mas,
which places it at a distance d$\approx32$pc. The star was classified
as K4 by G. P. Kuiper (published posthumously by Bidelman 1985), and
listed as K3 in the catalog of {\it Nearby Stars, Preliminary 3rd
Version} (Gliese \& Jahreiss 1991). In 1991, the most accurate
parallax measurements placed the star just within the 25pc limit of
the Gliese \& Jahreiss catalog, although the star is now believed to
be outside the $25$pc range.

We have found an $r=16.0$ star $40.8\arcsec$ to the southwest of GJ
4047 which has a proper motion extremely similar to GJ 4047 (Figure
1). On the POSS-I plate, the object is blended with a background
star. Our Lick spectrum shows the star to be an M5 V dwarf, which is
largely consistent with the star being at 32pc. We conclude that the
star is a distant companion on a long period orbit, with a current
projected separation r$\approx$1260 AU. We hereby adopt the name GJ
4047A for the primary, and GJ 4047B for the low-mass secondary.

We do measure a small relative proper motion of $\approx15$mas
yr$^{-1}$ between the two stars. The mean error on our proper
motion measurements is $\approx 5$ mas yr$^{-1}$. The measured
relative proper motion is thus significant at the $3\sigma$
level. Assuming a total system mass $\approx0.8M_{\sun}$, the orbital
period will be $\approx5\times10^{4}$ yr, which translates into a
possible relative proper motion $\approx 5$ mas yr$^{-1}$ at 32pc for
this system. Orbital motion is thus a reasonable explanation for the
observed relative proper motion.

\begin{figure}
\plotone{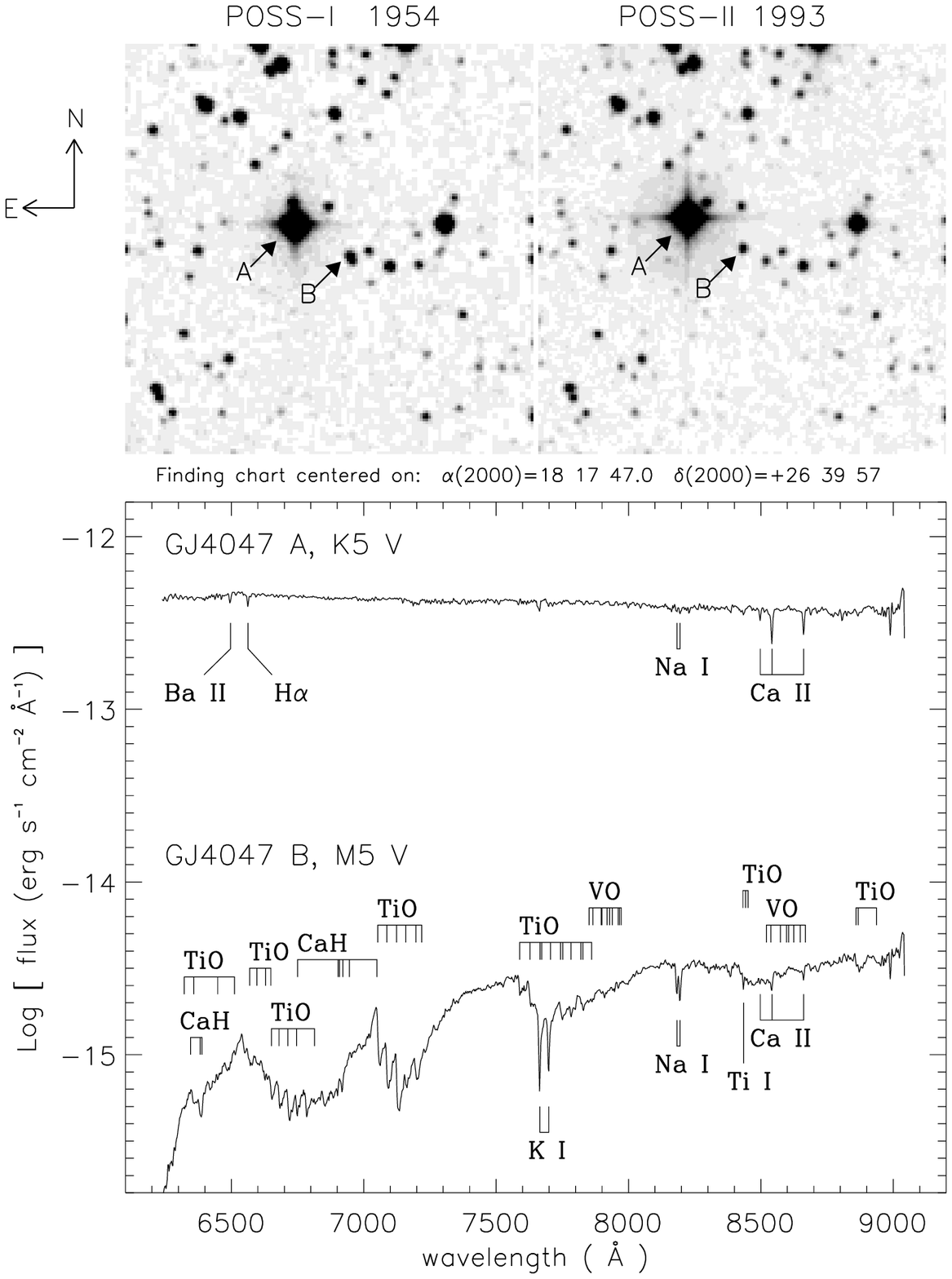}
\caption{\label{fig1} Top: finding chart for the star GJ 4047A and its
newly discovered distant companion GJ 4047B. Bottom: spectra of the
two stars plotted on on the same logarithmic scale. Atomic absorption
lines and main molecular bands are identified.}
\end{figure}

\placefigure{fig1}

\subsection{{\it GJ 718AB}}

GJ 718 is the bright variable star V774 Her, a known nearby flare
star. It is the HIPPARCOS star HIPP-90959, a V=8.90 star with proper
motion $\mu=505.2$ mas yr$^{-1}$, and parallax $\pi=42.67\pm1.26$ mas
which places it at a distance d$\approx$23pc. The star was classified
K4 V by Cowley, Hiltner \& Witt (1967), and K5 V by Stephenson (1986).

We have found an $r=14.9$ star $51.6\arcsec$ to the northwest which
shares the same proper motion as GJ 718 (Figure 2). The star is
blended with a background source in the POSS-I image. Our Lick
spectrum reveals a spectral type M4.5 V, consistent with the star
being a nearby low-mass dwarf. The star must be a companion on a long
period orbit, with a current projected separation $\approx$1190AU. We
hereby adopt the name GJ 718A for the primary, and GJ 718B for the
low-mass secondary. We do not find any significant relative proper
motion between the two stars.

\begin{figure}
\plotone{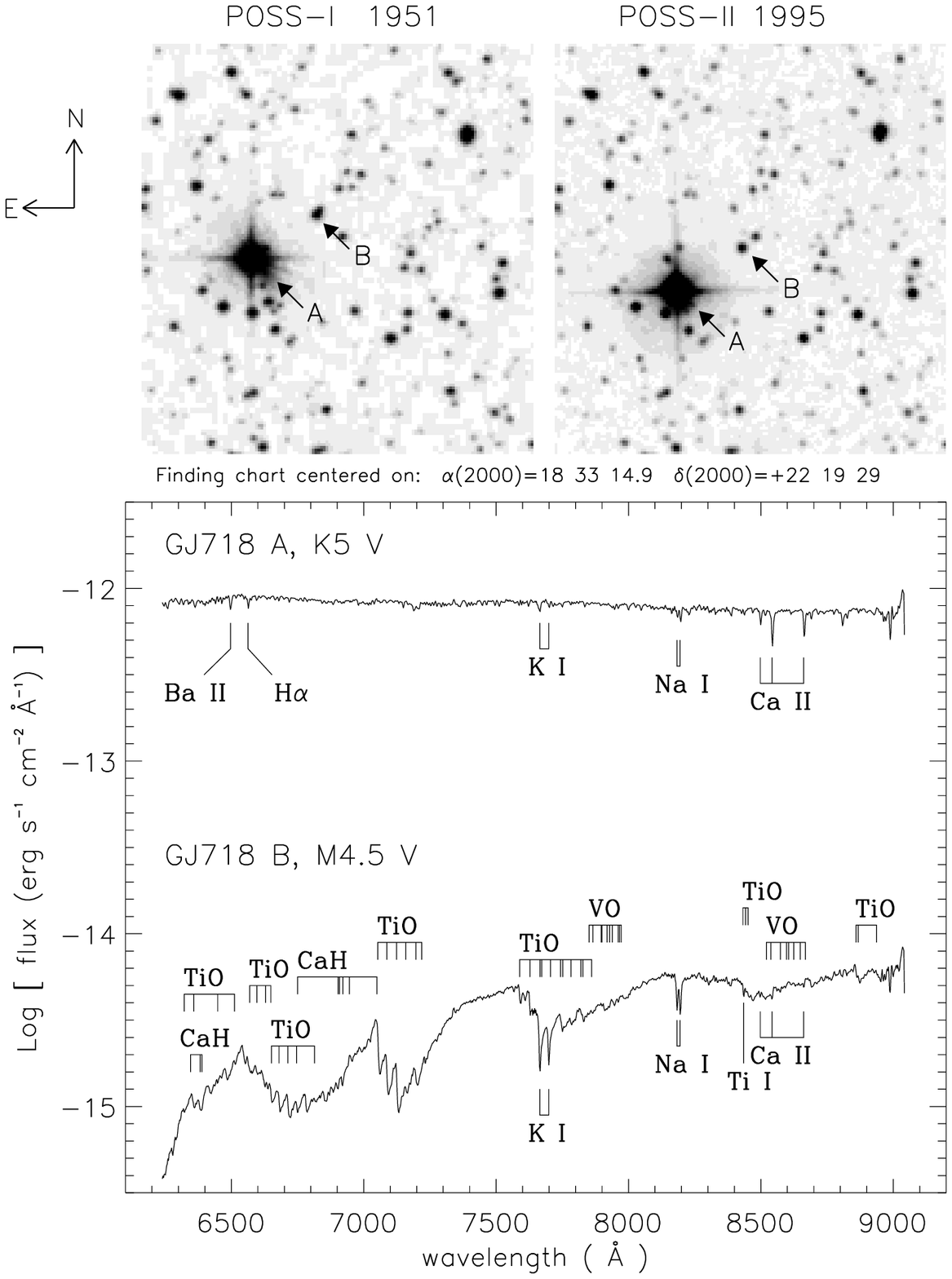}
\caption{\label{fig2} Finding chart and spectra for the GJ 718AB
system, as is Figure1.}
\end{figure}

\placefigure{fig2}

\subsection{{\it GJ 747.2(AB)C}}

The star GJ 747.2 is the double star COU 1462, resolved into two
components of equal magnitude by Couteau (1977). Speckle
interferometry measurements carried out in 1994 show two stars with a
separation $\rho=0.142\arcsec$ (Hartkopf {\it et al.} 2000). GJ 747.2
was observed as a single star by HIPPARCOS, and is now listed as a
V=9.42 star with parallax $\pi=34.37\pm1.30$ mas (catalog number HIP
94056), which places it at a distance of $\approx29$ parsecs. At this
distance, the projected separation between the close components is
only $\sim4$ AU, and the orbital motion should be observed on
timescales of years to decades. Observations carried out since the
1960s indeed show significant variations in both separation and
position angle (Couteau 1995), which can be attributed to the
orbital motion, although the orbit has not been modeled yet. We will
refer to the two components as GJ 747.2A and GJ 747.2B, though we
refrain from specifically assigning those names to one star or the
other at this point. This is meant only to identify the third member
of the system as GJ 747.2C.

The star we now name GJ 747.2C is an $r=13.6$ star lying $44.5\arcsec$
to the northeast of GJ 747.2AB, and having virtually the same proper
motion as the close pair (Figure 3). Our Lick spectra show the star to
be of spectral type M4 Ve, with the H$_{\alpha}$ line clearly in
emission. Assuming a distance of 29pc, the tertiary has a projected
separation of r$\approx$1290 AU. We measure a relative proper motion
between AB and C of $\approx10$ mas yr$^{-1}$, which is only
significant at the $2\sigma$ level but could be related to the orbital
motion of the tertiary. This system is listed as two point sources in
the {\it 2MASS Point Source Catalogue, 2MASS 2000 Second Incremental
Release}. Components A and B have a combined infrared magnitude
$K_s$=6.32, while component C has $K_s$=9.59.

\begin{figure}
\plotone{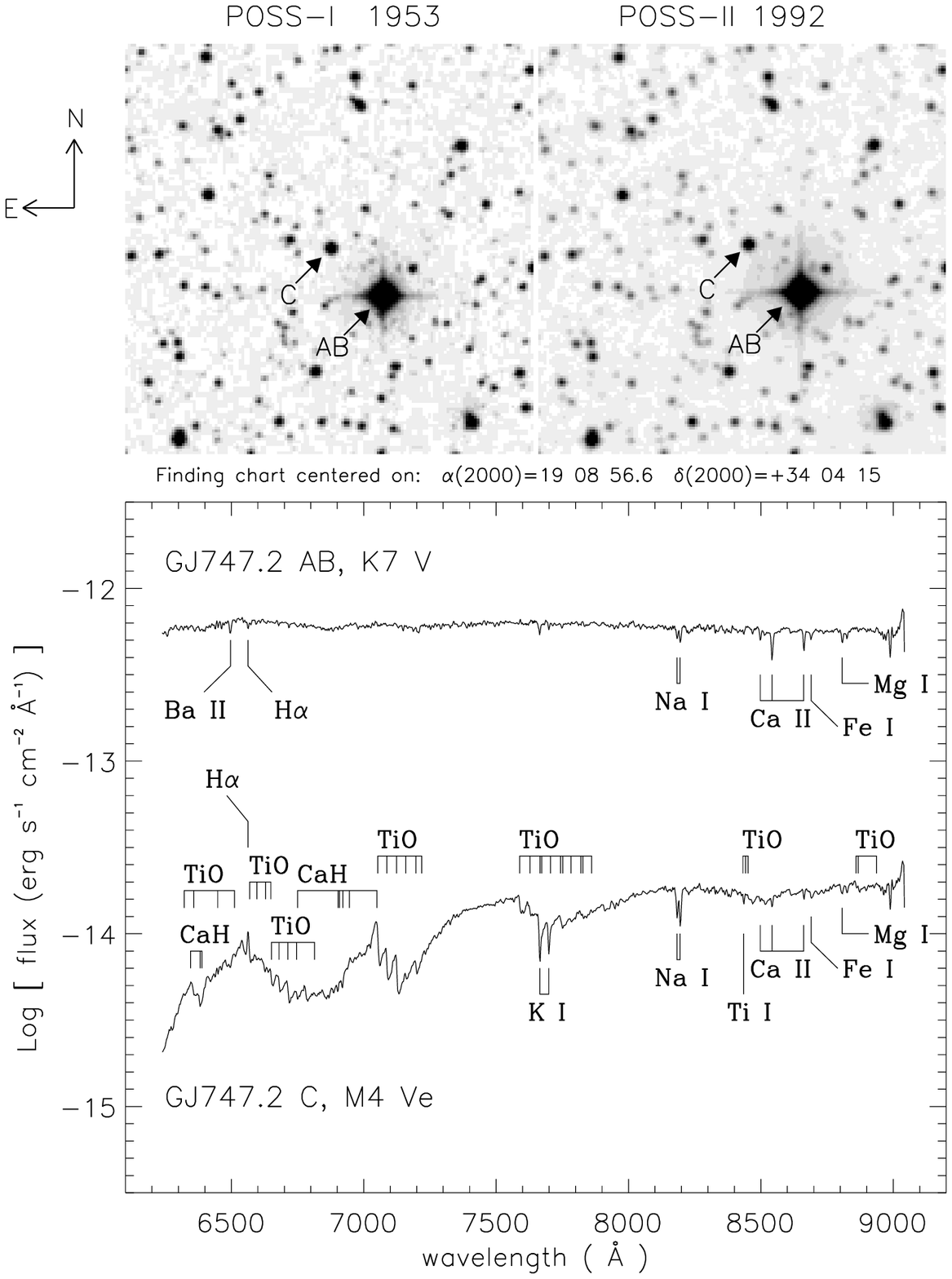}
\caption{\label{fig3} Finding chart and spectra for the GJ 747.2ABC
system, as is Figure1. Components A and B are only $0.142\arcsec$ apart,
and are not resolved on the charts. A and B and not resolved
spectroscopically either. Since both stars have approximately the same
magnitude, it is assumed that they both have a K7 V spectral type.}
\end{figure}

\placefigure{fig3}

\subsection{{\it GJ 4100AB}}

This low galactic latitude (b=-3.57$^{\circ}$) star was found in the
Lowell proper motion survey (Giclas, Slaughter \& Burnahm
1959). While it has no measured trigonometric parallax, its color and
spectral type makes it a good candidate solar neighborhood star; it
was included in the {\it Preliminary Version of the third catalogue of
nearby stars} (Gliese \& Jahreiss 1991). Its estimated distance is
15pc$\lesssim d \lesssim$45pc. This V$\simeq$12th magnitude star was
not in the HIPPARCOS input catalog.

In the very crowded field around GJ 4100, we have found an
$r=16.6$th magnitude star $13.4\arcsec$ to the northwest of GJ
4100 (Figure 4). The star is moving in the same direction and at the
same proper motion rate as GJ 4100. Our Lick spectrum shows the star
to be a low-mass M dwarf of spectral type M4.5 V. We conclude that the
star is a companion to GJ 4100; we therefore rename GJ 4100 as GJ
4100A, and adopt the name GJ 4100B to the low-mass companion. 

We find a relative proper motion between GJ4100A and GJ4100B of
$\approx11$ mas yr$^{-1}$, significant only at the $2\sigma$ level. At
$\approx 45$pc, the outer limit of the Gliese \& Jahreiss (1991)
estimated distance range, the projected separation between the
components would be r$\approx 600$AU, giving an orbital period
$\approx 1.5\times10^4$ years. At 45pc, the expected orbital motion
would be on the order of 5 mas yr$^{-1}$, which would be consistent
with the marginal relative proper motion we measure.

The spectrum of GJ4100 B, when compared to the other M4.5 V star
GJ718B (see figure 1), shows what appears to be a weak excess emission
at the wavelength of the H$_{\alpha}$ line. A division between the
two spectra does indeed show an excess in H$_{\alpha}$ at $5\sigma$ of
the instrumental noise. Hence, GJ4100B is apparently an emission line
star, although the H$_{\alpha}$ emission is very weak and would
certainly have been missed on a noisier spectrum. We assign this star
a spectral type M4.5 Ve.

\begin{figure}
\plotone{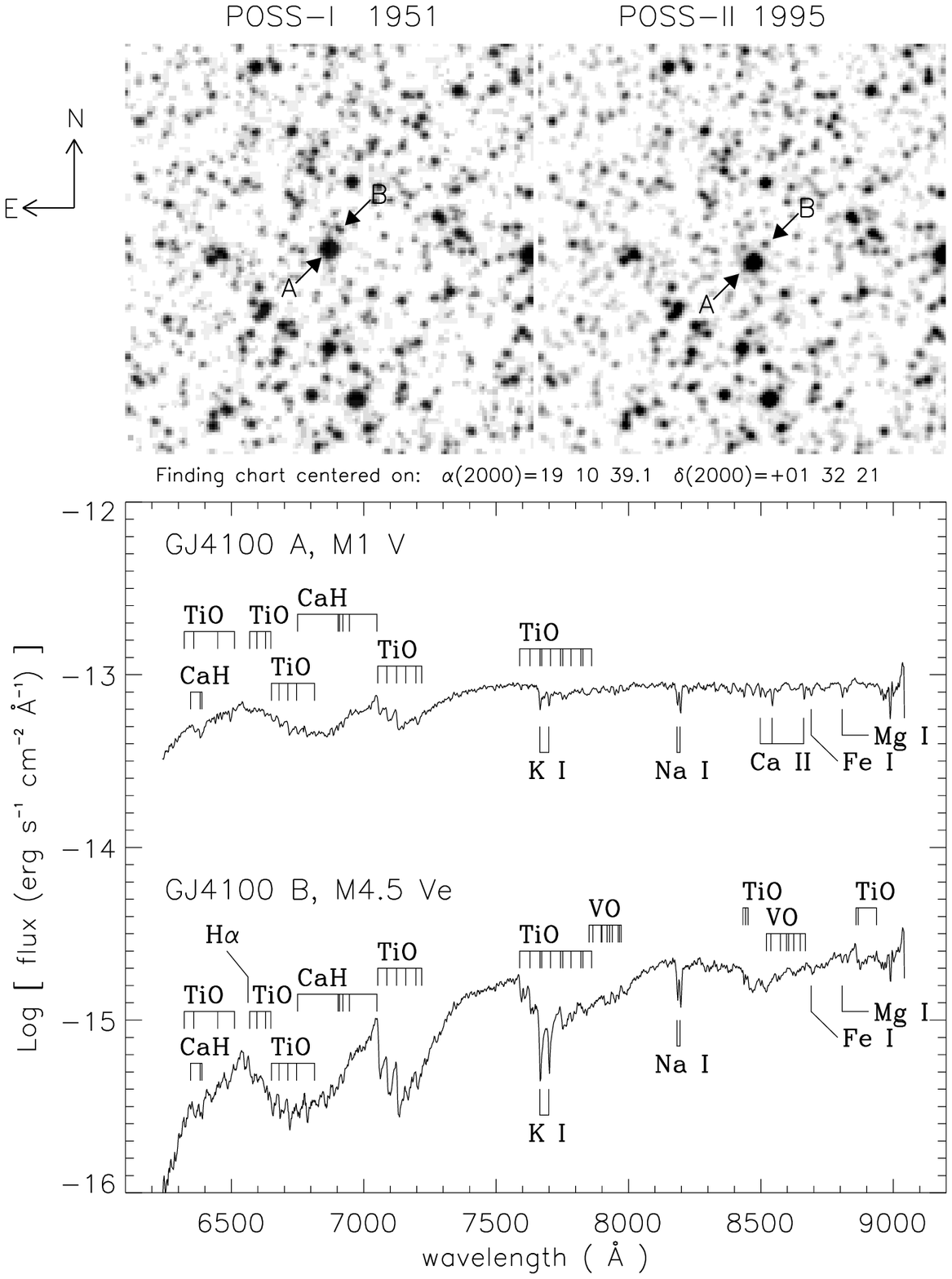}
\caption{\label{fig4} Finding chart and spectra for the GJ 4100AB
system, as is Figure1.}
\end{figure}

\placefigure{fig4}

\subsection{{\it GJ 4153AB}}

This is the high proper motion star Wolf 1351, a known flare star. No
trigonometric parallax measurement exists for this star, but Gliese \&
Jahreiss (1991) give an estimated parallax px=43.0$\pm$7.0 mas, based
on the color and spectral type. The star is thus estimated to be in
the distance range 20pc$\lesssim d \lesssim 28$pc.

The $r=14.7$ proper motion companion $50.5\arcsec$ to the southeast
(Figure 5) was serendipitously discovered by McCarthy, Zuckermann, \&
Becklin (2001), in a search for close companions to nearby
stars. Based on optical and infrared colors, they suggested a spectral
type M2.5 V. Our Lick spectrum shows the star to be M3.5 V. We assign
the name GJ 4153A to the primary, and GJ 4153B to its distant
companion. Both stars have counterparts in the {\it 2MASS Point Source
Catalogue, 2MASS 2000 Second Incremental Release}, with infra-red
magnitudes $K_s$=7.36 and 10.60, respectively; McCarthy, Zuckerman, \&
Becklin (2001) have measured similar values of $K$=7.39 and 10.60.

\begin{figure}
\plotone{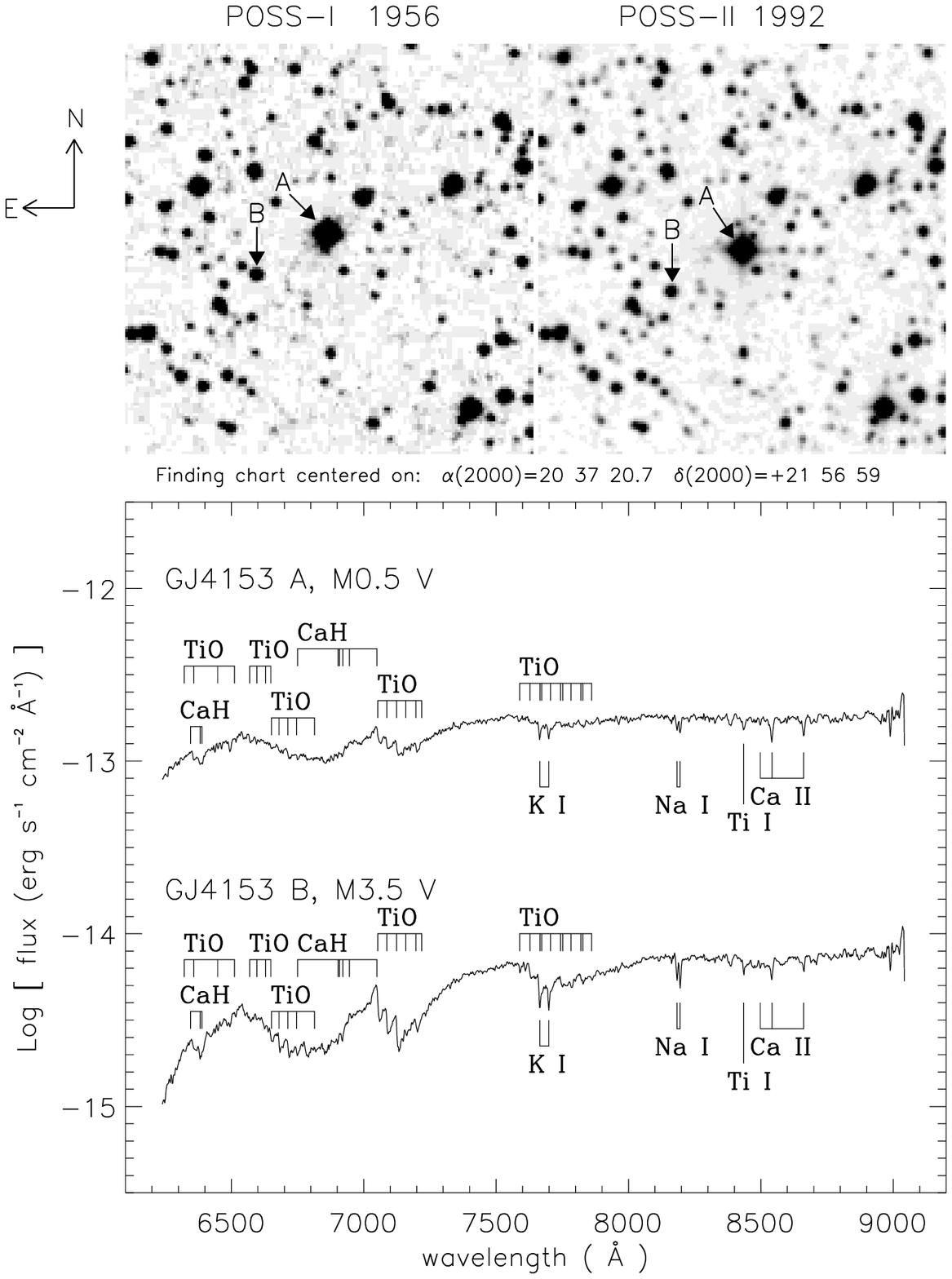}
\caption{\label{fig5} Finding chart and spectra for the GJ 4153AB
system, as is Figure1.}
\end{figure}

\placefigure{fig5}

\section{Conclusions}

We have confirmed the suspicion of McCarthy, Zuckerman, \& Becklin
(2001) that a significant number of known nearby stars have
unrecognized distant companions that are recorded in the Digitized Sky
Survey. We believe that this is largely due to the incompleteness of
published proper motion surveys, which is especially bad in the
crowded fields of the Milky Way. Our automated search for high proper
motion stars at low galactic latitudes ($-25^{\circ}<b<+25^{\circ}$)
in the region $16h<$RA$<22h$ DEC$>-2.5^{\circ}$, has turned up over
2000 new high proper motion stars, of which 4 were found to be
unrecognized common proper motion companions of known nearby stars (we
also recovered the companion found by McCarthy, Zuckerman, \& Becklin
2001).

Our survey area contained only 346 of the stars listed in the {\it
Preliminary Version of the Third Catalogue of Nearby Stars} (Gliese \&
Jahreiss 1991), which lists a total of 3803 stars. A simple
extrapolation suggests that there exist $\approx$50 more unrecognized
common proper motion companions to GJ stars. However, our survey
covers mainly crowded, low galactic latitude fields where we do expect
to find a larger number of unidentified (especially faint) proper
motion companions. It is likely that fewer common proper motion
companions to GJ star remain to be found at higher galactic
latitudes. Nevertheless, our survey has so far covered only 25\% of
the low galactic latitude fields; it is thus reasonable to expect that
at least $\approx15$ more unrecognized proper motion companions will
be found.

On the other hand, there are known wide binaries with angular
separations larger than the 1.5$\arcmin$ limit used in the present study.
Out of the 224 multiple systems with separation $\rho>10\arcsec$
listed in Poveda {\it et al.} (1994), 52 have a separation
$\rho>1.5\arcmin$. Given that common proper motion doubles are easier
to identify when they are relatively close to each other, we strongly
suspect that there are still more unidentified wide binaries with
$\rho>1.5\arcmin$. We therefore plan to extend our search for wide binaries
to potential proper motion doubles with larger angular separations.

\acknowledgments

This research program is supported by NSF grant AST-0087313 at the
American Museum of Natural History. This research has made use of the
SIMBAD astronomical database and VizieR catalogue access tool,
operated by the Centre de Donn\'ees Astronomiques de Strasbourg
(http://cdsweb.u-strasbg.fr/), and of the NStars database, operated by the
Nearby Stars Database Project at NASA-Ames research center
(http://nstars.arc.nasa.gov/). We are extremely grateful to Brian
McLean, director of the Catalogs and Surveys Branch at the Space
Telescope Science Institute, and his team, for their support in
gaining access to the Digitized Sky Survey database.

\newpage

\newpage

\begin{deluxetable}{lrrrrrrrr}
\tabletypesize{\footnotesize}
\tablecolumns{9} 
\tablewidth{0pc} 
\tablecaption{New nearby doubles: coordinates, proper motions, magnitudes} 
\tablehead{
\colhead{Star} & 
\colhead{$\alpha$(2000.0)\tablenotemark{1}} &
\colhead{$\delta$(2000.0)\tablenotemark{1}} &
\colhead{$\dot{\alpha}$(mas yr$^{-1}$)} & 
\colhead{$\dot{\delta}$(mas yr$^{-1}$)} &
\colhead{$b$}\tablenotemark{2} &
\colhead{$v$}\tablenotemark{3} &
\colhead{$r$}\tablenotemark{4} &
\colhead{$K_s$}\tablenotemark{5}
}
\startdata 
GJ4047A  & 18 17 49.83 & +26 40 17.8 & +326 & +112 & 10.9   &   9.7   & \nodata & \nodata \\ 
GJ4047B  & 18 17 47.19 & +26 39 57.5 & +311 & +116 & 18.3   & \nodata &  16.0   & \nodata \\ 
GJ718A   & 18 33 17.74 & +22 18 46.5 & -176 & -473 & 10.4   &   9.0   & \nodata & \nodata \\ 
GJ718B   & 18 33 14.76 & +22 19 17.3 & -172 & -472 & 17.6   & \nodata &  14.9   & \nodata \\ 
GJ747.2AB& 19 08 53.87 & +34 03 44.6 &  -82 &  +54 & 11.2   &   9.6   & \nodata &  6.32   \\ 
GJ747.2C & 19 08 56.53 & +34 04 14.4 &  -88 &  +63 & 15.9   & \nodata &  13.6   &  9.59   \\ 
GJ4100A  & 19 10 38.54 &  +1 32 10.4 & -167 & -192 & \nodata & 12.0   & \nodata & \nodata \\ 
GJ4100B  & 19 10 38.05 &  +1 32 21.7 & -159 & -200 & 19.0   & \nodata &  16.6   & \nodata \\ 
GJ4153A  & 20 37 20.77 & +21 56 48.8 &  -45 & -296 & 12.7   &  11.2   & \nodata &  7.36   \\ 
GJ4153B  & 20 37 24.03 & +21 56 26.3 &  -40 & -294 & 17.1   & \nodata &  14.7   & 10.60   \\ 
\enddata 
\tablenotetext{1}{J2000 equatorial coordinates at epoch 2000.00}
\tablenotetext{2}{POSS-II J blue magnitudes (IIIaJ + GG385)}
\tablenotetext{3}{Pal Q-V visual magnitude (IIaD + W12)}
\tablenotetext{4}{POSS-II F red magnitude (IIIaF + RG610)}
\tablenotetext{5}{2MASS infrared magnitude}
\end{deluxetable} 

\begin{deluxetable}{lrrrrl}
\tablecolumns{5} 
\tablewidth{0pc} 
\tablecaption{New nearby doubles: spectroscopic indices, spectral type} 
\tablehead{
\colhead{Star} & 
\colhead{CaH1} &
\colhead{CaH2} &
\colhead{CaH3} &
\colhead{TiO5} &
\colhead{spectral type}
}
\startdata 
GJ4047A  & 0.992 & 0.998 & 0.987 & 0.982 &  K5 V   \\ 
GJ4047B  & 0.703 & 0.306 & 0.594 & 0.271 &  M5 V   \\ %16.00->40pc

GJ718A   & 0.995 & 0.995 & 0.985 & 0.981 &  K5 V   \\ 
GJ718B   & 0.790 & 0.357 & 0.654 & 0.315 &  M4.5 V \\ %14.87->35pc

GJ747.2AB& 0.955 & 0.926 & 0.953 & 0.935 &  K7 V   \\ %11.17->30pc
GJ747.2C & 0.735 & 0.391 & 0.638 & 0.407 &  M4 Ve  \\ %13.60->35pc

GJ4100A  & 0.835 & 0.610 & 0.812 & 0.657 &  M1 V   \\ %
GJ4100B  & 0.817 & 0.342 & 0.657 & 0.316 &  M4.5 Ve \\ %16.58->80pc

GJ4153A  & 0.846 & 0.660 & 0.819 & 0.700 &  M0.5 V \\ %12.65->35pc
GJ4153B  & 0.783 & 0.440 & 0.708 & 0.446 &  M3.5 V \\ %17.06->80pc
\enddata 
\end{deluxetable} 

\end{document}